\documentclass[conference]{IEEEtran}
\IEEEoverridecommandlockouts
\usepackage{cite}
\usepackage{graphicx} 
\usepackage{xurl}
\usepackage{amsmath,amssymb,amsfonts}
\usepackage{textcomp}
\usepackage{hyperref}
\usepackage{listings}
\usepackage{xcolor}

\usepackage{tabularx}
\usepackage{booktabs}
\usepackage{array}
\usepackage{subcaption}
\usepackage{makecell}
\usepackage{siunitx}
\usepackage{enumitem}
\usepackage{adjustbox}

\newcommand{\spr}{\mathrm{SPR}}
\newcommand{\aonetwo}{\hat{A}_{12}}
\newcommand{\gdelta}{\Delta G_{2q}}
\newcolumntype{Y}{>{\raggedright\arraybackslash}X}

\def\BibTeX{{\rm B\kern-.05em{\sc i\kern-.025em b}\kern-.08em
    T\kern-.1667em\lower.7ex\hbox{E}\kern-.125emX}}

\begin{document}
\title{Non-Functional Requirements to Hyperparameter Translation for Quantum Software via LLMs}
\author{\IEEEauthorblockN{Joseph G. Bronyah}
\IEEEauthorblockA{\textit{University of Tennessee} \\
Knoxville, TN, USA \\
jbronyah@vols.utk.edu}
\and
\IEEEauthorblockN{Nasir U. Eisty}
\IEEEauthorblockA{\textit{University of Tennessee} \\
Knoxville, TN, USA \\
neisty@utk.edu}
}

\lstset{
    frame=single, 
    basicstyle=\ttfamily\footnotesize, 
    keywordstyle=\color{blue}, 
    commentstyle=\color{gray}, 
    stringstyle=\color{red}, 
    breaklines=true, 
    showstringspaces=false, 
    captionpos=b, 
    escapeinside={(*@}{@*)}, 
}

\maketitle

\begin{abstract}
Quantum software engineering (QSE) lacks validated middleware connecting natural-language non-functional requirements (NFRs) to low-level execution parameters such as shots and transpiler optimization level; existing translation approaches are judged on text similarity rather than execution consequences.
We test whether large language models (LLMs) can serve as that middleware layer, translating NFRs into Qiskit configurations whose downstream time, cost, and sampling behavior we then measure directly against a deterministic heuristic baseline.
We compare seven LLM agents with the heuristic on 540 synthetic and 18 real-world user stories, simulating predicted and ideal configurations on Qiskit/Aer circuits transpiled for \texttt{FakeBrisbane}. We report semantic pass rate, time, cost, top-state probability, and two-qubit gate exposure, using exact binomial intervals, Cochran's Q, McNemar, Friedman/Wilcoxon tests, Vargha--Delaney effect sizes, and bootstrap domain-shift analysis.
Every LLM significantly outperformed the heuristic on semantic pass rate after correction (real-world rates 55.6--83.3\% versus 16.7\%); only the heuristic showed detectable domain shift. Pairwise LLM differences did not survive correction on the real-world holdout, so no single model is established as superior. Speed/cost configurations were consistently faster and cheaper, but top-state probability did not support the fidelity hypothesis. A subset of apparent failures was instead traced to provider safety refusals on hazard-themed stories, conflating translation error with moderation policy.
LLM-based NFR-to-configuration translation, validated against execution consequences rather than text similarity, is a viable middleware layer for QSE when paired with simulation-based checks and human oversight, though cross-model ranking claims should remain cautious and refusal effects should be disentangled from translation quality.
\end{abstract}

\begin{IEEEkeywords}
Quantum software engineering, large language models, requirements engineering, non-functional requirements
\end{IEEEkeywords}

\section{Introduction}\label{sec:introduction}

Requirements engineering (RE) has long grappled with the gap between how stakeholders express intent and what a system must be configured to do. NFRs such as ``respond quickly,'' ``minimize cost,'' or ``prioritize accuracy'' are typically stated qualitatively, while the systems that must satisfy them expose only quantitative, low-level configuration surfaces. Closing this gap traditionally requires a human analyst, or increasingly, an automated mediator capable of interpreting natural language and producing valid configuration artifacts. The risk is not merely cosmetic: a mediator that misreads intent silently produces a system that runs but does not honor the stakeholder's actual priorities, an operational failure that conventional schema validation cannot detect because the output remains syntactically well formed.

QSE offers a setting where this RE problem is unusually sharp and unusually measurable. It is sharp because quantum execution frameworks expose few but highly consequential configuration parameters; in Qiskit, developers must set the number of circuit shots and a transpiler \texttt{optimization\_level}, where IBM's documentation defines levels from 0 (minimal optimization) to 3 (most aggressive)~\cite{ibmqiskit2026,qiskitoptimizationlevels2026,qiskittranspilerapi2026}. It is measurable because a misconfigured choice has a quantifiable downstream cost: a developer who over-provisions shots for a latency-critical workload wastes time and money; a developer who under-provisions shots for a precision-sensitive workload degrades sampling reliability; and aggressive transpiler optimization that helps in one context can be inappropriate overhead in another. 

Unlike many classical configuration-translation problems, where the cost of a wrong choice is diffuse or delayed, the consequence here can be simulated and quantified directly. QSE is therefore not just an application domain for this study; it is a tractable instrument for studying NFR-to-configuration translation as a measurable RE phenomenon. The Talavera Manifesto frames QSE as an engineering discipline requiring explicit principles and commitments~\cite{piattini2020talavera}, and recent mapping studies show that requirements and operational configuration remain comparatively underdeveloped relative to testing and implementation~\cite{destefano2023quantumfrontier}.

LLMs offer a possible mediator because they can interpret natural-language context and generate structured outputs. Systematic reviews of LLMs for software engineering show growing use of LLMs across code, requirements, testing, repair, and documentation tasks, while also emphasizing risks such as hallucination and the need for rigorous, task-specific evaluation~\cite{hou2024llmse}. This study does not propose or ship a deployable NFR-to-configuration tool. Instead, it proposes an evaluation framework to ask a narrower and more falsifiable RE question: can off-the-shelf LLMs serve as this translation mediator with enough semantic fidelity that, when their output is actually executed, the requirement's intent is preserved, not just approximated in text?
This paper makes four contributions:
\begin{enumerate}
    \item \textbf{Semantic translation framework:} We define a pipeline that translates natural-language user stories into Qiskit configuration parameters: \texttt{shots} and \texttt{optimization\_level}.
    \item \textbf{Two-tier benchmark:} We provide 540 synthetic stories and 18 real-world holdout stories for both controlled and external evaluation.
    \item \textbf{Operational simulation:} We connect semantic translation accuracy to execution consequences using Qiskit/Aer proxy circuits compiled against \texttt{FakeBrisbane}~\cite{fakebrisbane2026}.
    \item \textbf{Correctness-to-cost coupling:} Rather than scoring translations on textual or numeric similarity to a reference output, we measure whether semantic errors propagate into quantifiable execution penalties in time, cost, sampling behavior, and circuit structure, directly testing whether NFR violations have operational consequences and not only annotation disagreement.
\end{enumerate}

\section{Background}\label{sec:background}
\subsection{QSE and Requirements Translation}
QSE extends classical software engineering to systems involving quantum algorithms, quantum hardware constraints, hybrid workflows, and probabilistic outputs. Unlike many classical systems, the execution behavior of quantum programs is strongly influenced by configuration decisions made at runtime or compile time, including circuit transpilation, backend selection and measurement shots.

The practical gap addressed in this paper is that QSE lacks standardized middleware that translates high-level business goals into low-level execution parameters. A domain expert may know that a logistics optimizer must produce a result quickly during a supply-chain disruption, but may not know how many shots or which transpiler optimization level is appropriate. Conversely, a quantum developer may understand the execution parameters but misread the business priority embedded in the requirement.

\subsection{Quantum Configuration and Semantic Middleware}
This study focuses on two configuration parameters. The number of \texttt{shots} specifies repeated circuit executions used to sample measurement outcomes and build a statistically significant probability distribution. Higher values generally support more stable estimates but increase runtime and cost. The \texttt{optimization\_level} specifies Qiskit's transpiler optimization level from 0 to 3. Higher values generally spend more effort optimizing circuit structure for physical hardware. These two parameters form a compact but meaningful configuration space for evaluating semantic translation.

The Qiskit configuration surface includes other parameters, such as backend selection, error mitigation method, and qubit layout, that also affect execution behavior. We restrict the scope to \texttt{shots} and \texttt{optimization\_level} because they are exposed identically across virtually every Qiskit backend and algorithm family, whereas error mitigation method and qubit layout are backend- or hardware-specific and would confound model comparisons with hardware-availability effects rather than translation quality. Both chosen parameters also map cleanly onto a single underlying NFR axis, speed/cost versus accuracy/fidelity, which keeps the ground-truth rubric (Table~\ref{tab:rubric}) unambiguous. We treat this pair as a minimal but representative probe of NFR-to-configuration translation rather than a complete configuration-management solution; Section~\ref{sec:futurework} discusses extending the same evaluation methodology to a richer parameter space.

The central assumption is that LLMs can infer technical configuration intent from natural language. The translation agent receives a user story and returns a JSON object as:
\begin{verbatim}
{  "predicted_shots": 100,
  "predicted_optimization_level": 0 }
\end{verbatim}
The core prompts are reproduced in Appendix~\ref{app:prompts}. A separate LLM judge evaluates the configuration against an intent-aware rubric. The judge does not require exact equality because multiple parameter values may satisfy the same NFR category.

\section{Related Work} \label{sec:relatedwork}
\subsection{Quantum Software Engineering: Foundations, Lifecycle, and Requirements}
QSE has moved from a conceptual proposal toward an active research area. The Talavera Manifesto argued for systematic engineering principles over purely physics-driven experimentation~\cite{piattini2020talavera}, and subsequent mapping studies confirm rapid but uneven growth: testing dominates, while management, requirements engineering, and lifecycle support remain comparatively underdeveloped~\cite{destefano2023quantumfrontier,murillo2025roadmap}. Platform- and lifecycle-focused reviews extend this picture, surveying components, platforms, and development-lifecycle models~\cite{serrano2022components,dwivedi2024survey,desdentado2025mapping}, proposing end-to-end lifecycles from problem formulation through execution and analysis~\cite{weder2020quantumsoftwarelifecycle}, and examining quantum-computing-as-a-service and DevOps-style hybrid pipelines~\cite{ahmad2023qcaas,khan2025advancing}. Model-driven approaches push abstraction further by generating Qiskit code from UML or LLM/RAG-assisted models~\cite{jimenez2025umlcode,siavash2025model}. Collectively, these works are platform-, lifecycle-, or model-centric; our framework instead sits as middleware between requirements capture and executable configuration, translating NFRs directly into parameters rather than into code.

Requirements engineering for quantum software is a small but growing subfield. Existing studies argue that quantum requirements engineering differs fundamentally from its classical counterpart because quantum software is probabilistic, hybrid, resource-sensitive, and hardware-dependent, and identify recurring challenges in specifying requirements and quality attributes under these constraints~\cite{yue2023qsre,sepulveda2024requirements}. Complementary work on quantum quality attributes shows that platform characteristics and even mathematically equivalent problem formulations can yield substantially different maintainability, reliability, and performance outcomes~\cite{sodhi2021quantum,krueger2020quantum}. These studies motivate treating stakeholder language, e.g., ``fast,'' ``precise,'' ``low cost,'' as a first-class translation target; our work operationalizes this by treating NFR translation as a measurable semantic task linked to runtime, cost, and sampling penalties.

\subsection{Testing, LLMs, and Quantum Programming}
Testing and debugging have received the most empirical attention in QSE, motivated by the probabilistic, hard-to-observe nature of quantum programs. Large-scale bug studies and survey work show that a substantial share of failures are quantum-specific and that classical unit-testing intuitions transfer poorly~\cite{paltenghi2022bugs,Paltenghi2024TestingAnalysis,ramalho2025road,zappin2026challenges}, and static-analysis tools such as LintQ, together with benchmarks like QBugs and Bugs4Q, support this direction with concrete tooling~\cite{paltenghi2023lintq,campos2021qbugs,zhao2023bugs4q}. This line of work assumes a circuit already exists; we instead examine the earlier decision point where natural-language intent is converted into execution parameters before any program runs.

LLMs are increasingly studied as software engineering assistants, with systematic reviews emphasizing both their promise and the need for task-specific evaluation given risks such as hallucination~\cite{hou2024llmse}. Quantum-focused studies extend this lens: GPT-4 struggles to select hardware-appropriate operations and settings when writing circuits~\cite{henderson2025programming}, QuanBench reports frequent semantic errors across 44 quantum code-generation tasks~\cite{guo2025quanbench}, and PennyLang shows retrieval-augmented generation improves quantum code assistance accuracy~\cite{basit2025pennylang}. This literature focuses on generating or repairing code; we instead study whether LLMs preserve NFR intent when mapping requirements to the configuration parameters that govern how an already-correct program executes, a complementary evaluation layer for AI-assisted QSE.

\subsection{Classical NL-to-Configuration Translation and Positioning}
Translating natural-language intent into structured, executable configuration is not unique to quantum software. Text-to-SQL research tackles a closely related problem, with surveys documenting persistent ambiguity-resolution and schema-mapping challenges that mirror our NFR-to-category ambiguity~\cite{liu2025text2sql}. Infrastructure-as-code (IaC) generation is an even closer analogue, since correctness is defined by execution outcome rather than textual similarity: state-of-the-art LLMs achieve only 20.8--30.2\% first-attempt deployment success on a deployability-centric IaC benchmark~\cite{zhang2026deployability}, and follow-up work resolves underspecified IaC requests through hierarchical clarification questions rather than guessing~\cite{yang2026ambigiac}, an approach directly analogous to the human-clarification step we recommend in Section~\ref{sec:futurework}. These classical studies reinforce, outside quantum computing, this paper's central claim: an LLM's structured output can be semantically plausible without its execution preserving the original requirement's intent.

Together, this literature establishes that QSE is growing but still immature, that requirements and lifecycle support lag behind testing, that quantum software quality depends on execution configuration, and that LLMs are promising yet unreliable without task-specific validation, a pattern that recurs in classical NL-to-configuration tasks. This paper sits at the intersection of these themes: unlike platform surveys it does not catalog SDKs, unlike quantum testing studies it does not detect bugs in existing programs, unlike LLM code-generation benchmarks it does not ask whether a model can write a complete circuit, and unlike the classical NL-to-configuration literature it does not target SQL or cloud infrastructure. Instead, it evaluates whether LLMs preserve NFR intent when translating developer language into executable Qiskit configuration parameters, and whether semantic errors produce measurable operational penalties, applying execution-grounded evaluation logic from classical NL-to-configuration research to a quantum setting where it has not yet been tested.

\section{Methodology}\label{sec:methodology}
\subsection{Study Design}
\begin{figure}[!t]
    \centering
\includegraphics[width=0.82\columnwidth,height=0.58\textheight,keepaspectratio]{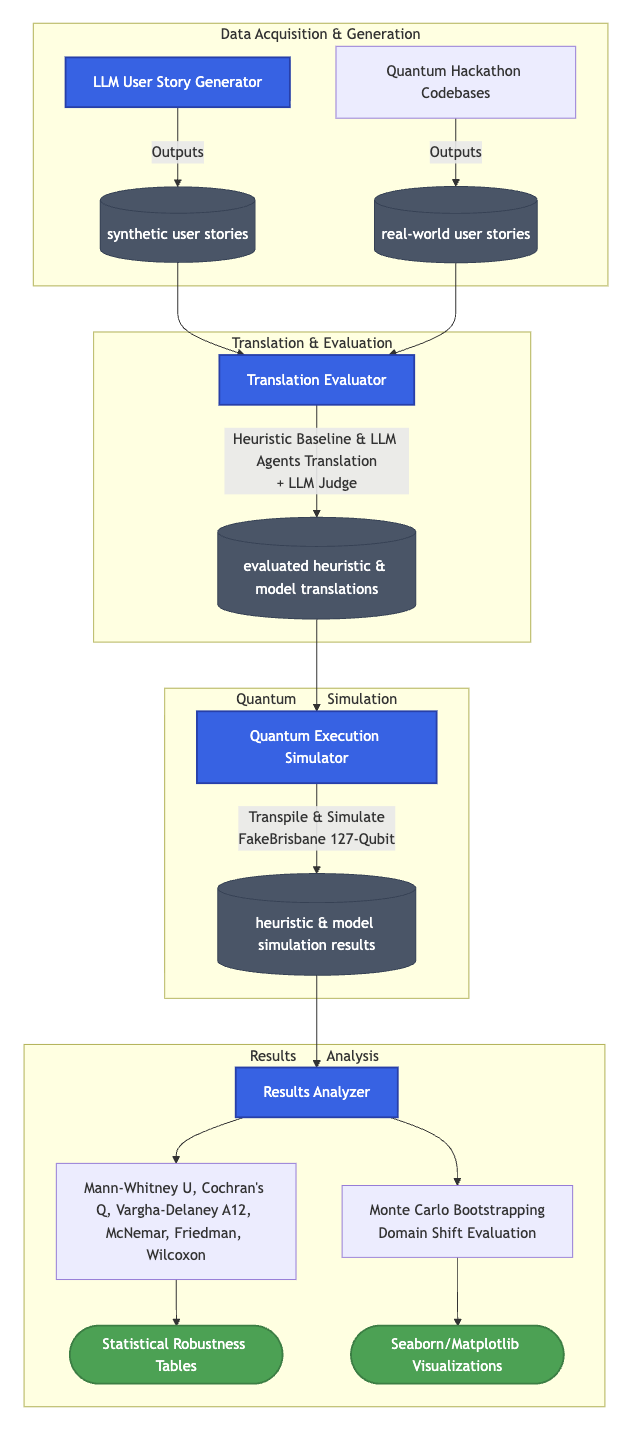}
    \caption{Implemented pipeline from dataset construction through translation, simulation, and statistical reporting.}
    \label{fig:methodflow}
\end{figure}

The study uses a controlled empirical design with two datasets, multiple translation agents, a deterministic baseline, simulation-based operational metrics, and statistical testing. The study follows four phases as summarized in Figure~\ref{fig:methodflow}.

\subsection{Research Questions and Hypotheses}
The study is organized around the three research questions listed in Table~\ref{tab:rqs}.

\begin{table}[h]
\caption{Research questions.}
\label{tab:rqs}
\begin{tabularx}{\linewidth}{@{}lY@{}}
\toprule
ID & Research question \\
\midrule
RQ1 & Do off-the-shelf LLMs translate natural-language NFRs into suitable quantum parameter categories more accurately than a deterministic heuristic? \\
RQ2 & How resilient are translation agents when moving from synthetic requirements to an independently curated real-world holdout? \\
RQ3 & What time, cost, sampling, and circuit-structure penalties arise when a translation is semantically misaligned? \\
\bottomrule
\end{tabularx}
\end{table}

The corresponding hypotheses are:
\begin{itemize}[leftmargin=*]
    \item \textbf{$H_{0,\mathrm{SPR}}$:} semantic pass probabilities are equal across the heuristic and LLM agents on the same stories. \textbf{$H_{A,\mathrm{SPR}}$:} at least one agent differs.
    \item \textbf{$H_{0,\mathrm{shift}}$:} a model's real-world pass rate is consistent with the pass-rate distribution obtained by drawing equally sized samples from its synthetic results. \textbf{$H_{A,\mathrm{shift}}$:} the real-world rate is inconsistent with that distribution.
    \item \textbf{$H_{0,\mathrm{time}}$:} among semantically successful translations, Speed/Cost and Accuracy/Fidelity stories have the same execution-time distribution. \textbf{$H_{A,\mathrm{time}}$:} Speed/Cost stories have lower execution time.
    \item \textbf{$H_{0,\mathrm{prob}}$:} the two intent groups have the same top-state probability distribution. \textbf{$H_{A,\mathrm{prob}}$:} Accuracy/Fidelity stories have higher top-state probability.
    \item \textbf{$H_{0,\mathrm{cost}}$:} the two intent groups have the same execution-cost distribution. \textbf{$H_{A,\mathrm{cost}}$:} Speed/Cost stories have lower cost.
    \item \textbf{$H_{0,G_{2q}}$:} absolute two-qubit gate deviation is distributed equally for semantic passes and failures. The directional exploratory alternative is that failures have greater deviation.
\end{itemize}

\subsection{Dataset}
The evaluation uses the two tiers summarized in Table~\ref{tab:dataset}: a synthetic benchmark of 540 stories and a real-world holdout set of 18 stories. The synthetic benchmark provides controlled coverage across QSE domains and NFR categories. The synthetic benchmark size is determined by the generator design rather than post-hoc selection. The \texttt{generate\_sample.py} script instantiates 54 domain–algorithm seed prompts, spanning 10 operational domains and 6 algorithm families. Each seed prompt is crossed with 10 intent phrasings: 4 focused on Speed/Cost, 3 Balanced, and 3 focused on Accuracy/Fidelity. This process creates a complete $54 \times 10$ matrix of 540 stories, all of which are evaluated exhaustively without subsampling. The real-world holdout contains 18 developer-style stories manually written from quantum hackathon codebases: for each codebase, an annotator inspected the implemented algorithm and any stated performance or accuracy goal and authored a user story describing the underlying intent in natural language, independent of the synthetic generation templates. A second author independently reviewed each story and its assigned ideal configuration (Table~\ref{tab:intents}) against the source codebase; disagreements were resolved by discussion before any translation or judging was performed. The per-story source codebases and annotation notes are included in the replication package (Section~\ref{sec:data}) rather than enumerated here.

Because the synthetic stories are themselves LLM-generated from author-authored templates, they may be unrealistically clean exemplars of their intended category, using more explicit or stereotyped speed/cost/accuracy language than an actual stakeholder would. This is a generator-bias threat distinct from judge bias: an overly clean synthetic benchmark could inflate measured pass rates relative to what naturally occurring requirements would produce. The real-world holdout, despite its small size, is the primary check on this threat, and the consistent (if uncertain) ranking between the two tiers is evidence against severe generator-bias inflation.

\begin{table}[h]
\caption{Two-tier dataset design.}
\label{tab:dataset}
\begin{tabularx}{\linewidth}{@{}l r Y Y@{}}
\toprule
Tier & Size & Purpose & Source/design \\
\midrule
Synthetic benchmark & 540 & Controlled evaluation across known domains and NFR categories & Generated from domain prompts and constraint mappings \\
Real-world holdout & 18 & Zero-shot domain-shift evaluation & Manually written from quantum hackathon codebases; sources archived in replication package\\
\bottomrule
\end{tabularx}
\end{table}

The synthetic generator maps constraint templates to ideal configurations. Speed/cost requirements map to low shots and low optimization; balanced requirements map to intermediate values; and accuracy/fidelity requirements map to high shots and high optimization. The synthetic user stories are distributed across ten distinct operational domains including Chemistry, Logistics, Finance, Cybersecurity and Material Science. Each domain prompt is tied to an algorithm family such as QAOA, Grover, QGAN, QSVM, VQE, or Hamiltonian Evolution.

\subsection{Intent Categories and Ground Truth}
Each story is assigned an ideal configuration using the categories in Table~\ref{tab:intents}.

\begin{table}[h]
\caption{Macro-intent categories and ideal configurations.}
\label{tab:intents}
\begin{tabularx}{\linewidth}{@{}Y l l Y@{}}
\toprule
Macro-intent & Example shots & Opt. level & Interpretation \\
\midrule
Speed/Cost Focus & 10, 100, 256 & 0--1 & Prioritize latency, cost, or rapid prototyping \\
Balanced Focus & 1024, 2048 & 1--2 & Balance execution speed with acceptable heuristic accuracy \\
Accuracy/Fidelity Focus & 8192, 10000 & 3 & Prioritize precision, reliability, or error mitigation \\
\bottomrule
\end{tabularx}
\end{table}

\subsection{Translation Agents and Baseline}
The study evaluates seven LLM-based translation agents and one deterministic baseline. The model labels are Qwen 80B, GPT OSS 120B, DeepSeek v4Pro, Qwen 235B, Gemma 4 31B, GPT 5.4, and Claude Sonnet 4.6. The deterministic baseline searches for speed and accuracy-related keywords and otherwise emits a balanced default. Both the baseline and LLM outputs are scored by the same rubric-driven judge. Prompt snapshots for dataset generation, translation, and judging are provided in Appendix~\ref{app:prompts}.

A prediction passes only when its shot category and optimization level both satisfy the rubric in Table~\ref{tab:rubric}.

\subsection{Semantic Pass Rubric}
A prediction is evaluated using a semantic pass/fail rubric rather than exact numeric equality. This is appropriate because multiple configurations may satisfy the same NFR category. Each criterion in Table~\ref{tab:rubric} is scored binarily by the judge (5 if satisfied, 0 otherwise; Appendix~\ref{app:prompts}), and a story passes only when both the shot and optimization criteria score 5.

\begin{table}[h]
\caption{Semantic pass criteria.}
\label{tab:rubric}
\begin{tabularx}{\linewidth}{@{}Y Y Y@{}}
\toprule
Intent type & Shot criterion & Optimization criterion \\
\midrule
Speed/Cost & Predicted shots $\leq 1000$ & Predicted optimization within $\pm 1$ of ideal \\
Accuracy/Fidelity & Predicted shots $\geq 2000$ & Predicted optimization within $\pm 1$ of ideal \\
Balanced/Scaling & Predicted shots between 500 and 4000 & Predicted optimization within $\pm 1$ of ideal \\
\bottomrule
\end{tabularx}
\end{table}

The primary translation metric is Semantic Pass Rate:
\begin{equation}
\spr_m = \frac{1}{N}\sum_{i=1}^{N}\mathbf{1}(\text{pass}_{m,i}=1),
\end{equation}
where $\spr_m$ is the semantic pass rate for model $m$, $N$ is the number of stories, and $\mathbf{1}$ is an indicator function.

\subsection{Quantum Simulation Pipeline}
For each story, the simulation pipeline constructs a proxy circuit corresponding to the recommended algorithm family. Table~\ref{tab:circuits} summarizes the mapping.

\begin{table}[h]
\caption{Proxy circuit mapping by algorithm family.}
\label{tab:circuits}
\begin{tabularx}{\linewidth}{@{}Y Y Y@{}}
\toprule
Algorithm family & Proxy circuit & Target use case \\
\midrule
QAOA & \texttt{QAOAAnsatz} & Optimization \\
VQE & \texttt{EfficientSU2} ansatz & Chemistry/materials simulation \\
QML/QSVM/QGAN & \texttt{ZZFeatureMap} + \texttt{RealAmplitudes} & Classification/generative learning \\
Grover & \texttt{GroverOperator} & Search \\
Hamiltonian Evolution & \texttt{PauliEvolutionGate} & Physical simulation \\
\bottomrule
\end{tabularx}
\end{table}

\begin{figure}[!t]
\centering
\begin{subfigure}[t]{0.48\columnwidth}
\centering
\includegraphics[width=\linewidth]{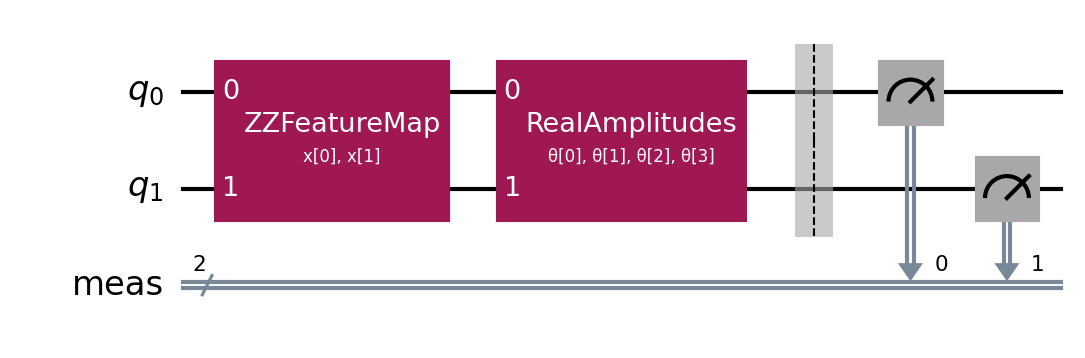}
\caption{QML proxy.}
\end{subfigure}\hfill
\begin{subfigure}[t]{0.48\columnwidth}
\centering
\includegraphics[width=\linewidth]{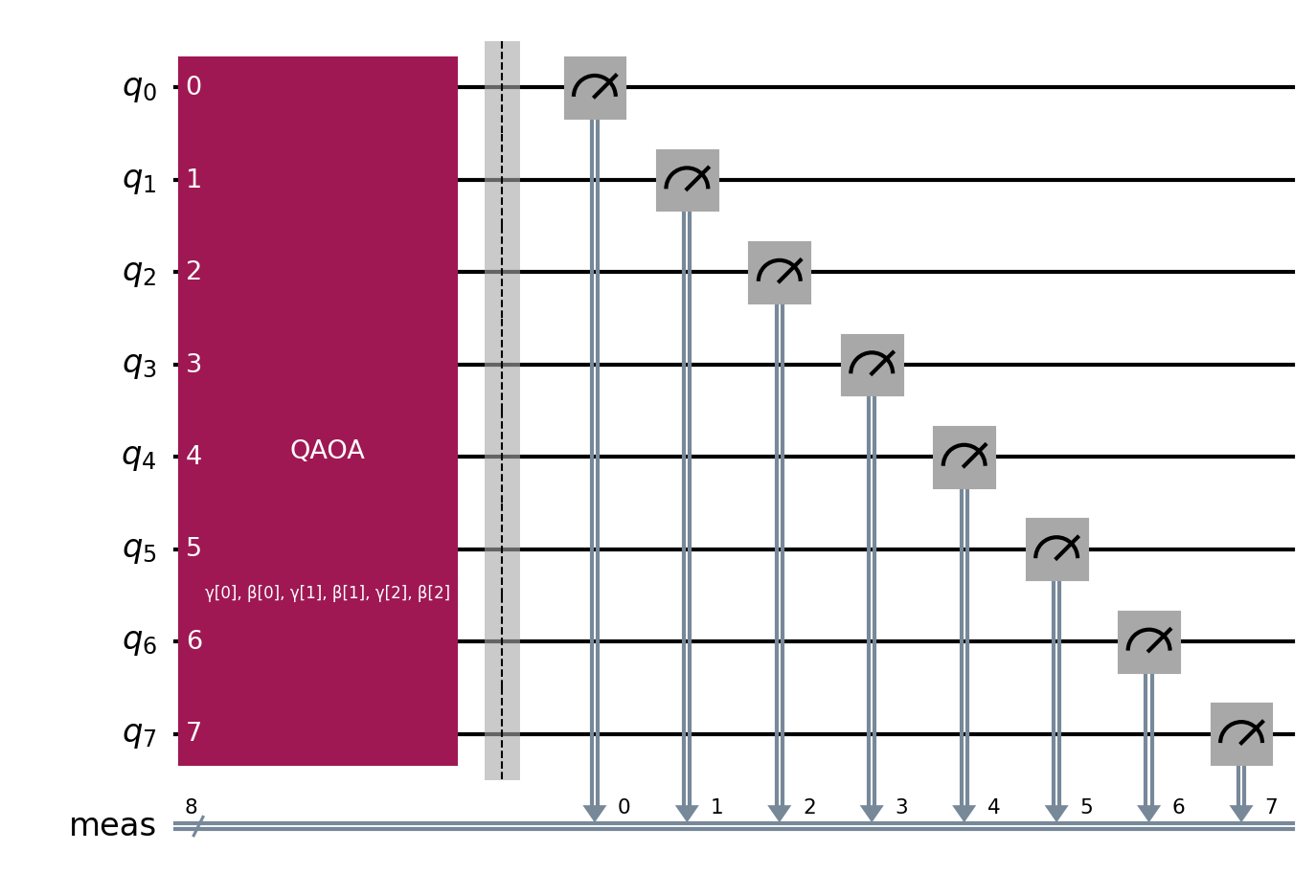}
\caption{QAOA optimization proxy, illustrated at eight qubits.}
\end{subfigure}

\vspace{0.35em}
\begin{subfigure}[t]{0.48\columnwidth}
\centering
\includegraphics[width=\linewidth]{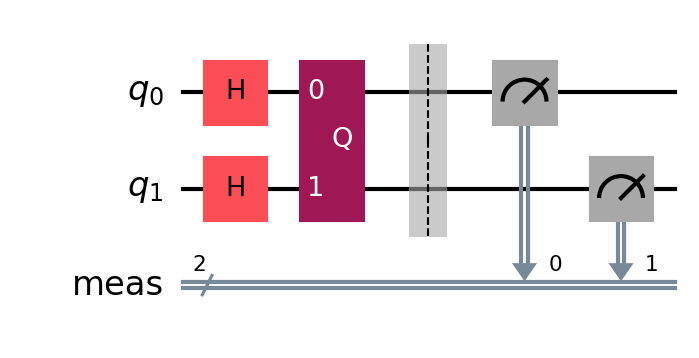}
\caption{Grover/search proxy.}
\end{subfigure}\hfill
\begin{subfigure}[t]{0.48\columnwidth}
\centering
\includegraphics[width=\linewidth]{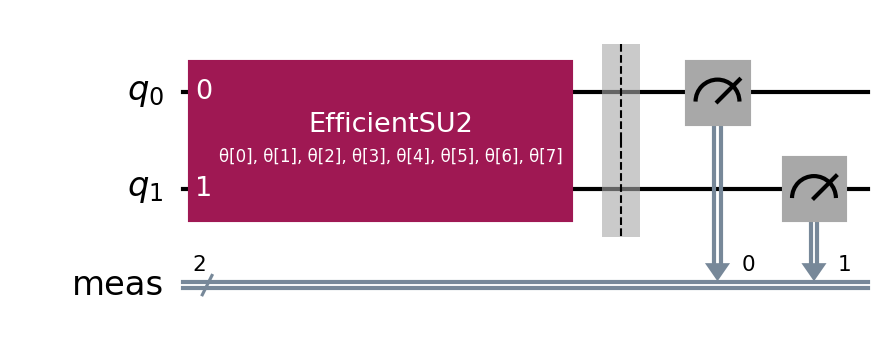}
\caption{VQE chemistry proxy.}
\end{subfigure}
\caption{Representative algorithm-family proxy circuits used by the simulation layer.}
\label{fig:circuits}
\end{figure}

The pipeline dynamically scales circuit width and depth based on complexity cues in the user story. Circuit width and repetitions are scaled from 2 qubits/1 repetition to 8 qubits/3 repetitions. Each circuit is transpiled for \texttt{FakeBrisbane} (a 127-qubit processor model) where predicted and ideal configurations are simulated independently. Figure~\ref{fig:circuits} shows representative pre-transpilation proxy circuits.

The execution-time is estimated as:
\begin{equation}
 t_{\text{shot}} = 35g_1 + 250g_2 + 5000m,
\end{equation}
\begin{equation}
 T_{\text{QPU}} = \text{shots} \times t_{\text{shot}} \times 10^{-9},
\end{equation}
where $g_1$ is the number of single-qubit gates, $g_2$ is the number of two-qubit gates, and $m$ is the number of measurements. The gate-time constants (35\,ns, 250\,ns, 5000\,ns) are illustrative superconducting-scale values, not calibration data from a specific backend; $T_{\text{QPU}}$ should therefore be read as a relative proxy for comparing predicted and ideal configurations on the same circuit, not as an absolute runtime prediction. For variational algorithms, per-iteration QPU time is multiplied by 50 iterations. Both variational algorithms and single-pass algorithms were executed with statistical sampling ($N=5$ per configuration) to isolate and average out stochastic compiler variance.
An illustrative rate of \$1.60 per QPU second, broadly representative of commercial cloud QPU pricing rather than a calibrated provider quote, was used to convert $T_{\text{QPU}}$ into a comparable cost proxy; absolute dollar values should not be interpreted as a specific vendor's billing.

\subsection{Operational Metrics}
For metric $M$, the configuration deviation is
\begin{equation}
\Delta M_i=M_{\mathrm{pred},i}-M_{\mathrm{ideal},i}.
\end{equation}
Primary operational deltas are time, cost, and top-state probability. Top-state probability is a sampling-concentration metric, not full statevector fidelity.

As a secondary hardware-exposure analysis, we compute
\begin{equation}
\gdelta=G_{2q}^{\mathrm{pred}}-G_{2q}^{\mathrm{ideal}},
\end{equation}
and its absolute value. Since entangling gates are important contributors to NISQ error exposure~\cite{preskill2018nisq,murali2019noise}, we also report a simple theoretical failure-rate calculation
\begin{equation}
R_{2q}=1-(1-\epsilon_{2q})^{G_{2q}},
\end{equation}
using $\epsilon_{2q}=0.01$. This quantity is not backend-calibrated fidelity but an approximate calculation of noise exposure.

\subsection{Statistical Analysis}
The analysis was pre-structured by dataset role. The 540-story synthetic benchmark supports primary inference. The 18-story real-world holdout is reported with descriptive summaries, exact tests, effect sizes, and bootstrap intervals, and its inferential findings are labeled exploratory.

For each model and dataset, we report Clopper--Pearson exact 95\% binomial confidence intervals~\cite{clopper1934confidence}. We use Cochran's Q omnibus test~\cite{cochran1950comparison} for pass/fail comparisons followed by pairwise exact McNemar tests~\cite{mcnemar1947sampling} with Holm correction~\cite{holm1979simple}. This is because all agents are evaluated on the same stories. Paired continuous outcomes use Friedman tests~\cite{friedman1937use} followed by Wilcoxon signed-rank tests \cite{wilcoxon1945individual} with Holm correction and paired rank-biserial effects. We use Holm rather than a Benjamini--Hochberg procedure because the pairwise comparisons feed claims about individual model-versus-baseline and model-versus-model differences, where a false positive is more costly than on a discovery-oriented analysis; Holm controls the family-wise error rate for these confirmatory comparisons, whereas Benjamini--Hochberg controls the (less conservative) false discovery rate more appropriate to exploratory screening across many simultaneous hypotheses.

Intent comparisons within successful synthetic translations use one-sided Mann--Whitney U tests and Vargha--Delaney $\aonetwo$~\cite{mann1947test,vargha2000critique}. Domain shift is assessed by drawing 5,000 samples of size 18 from each model's synthetic pass/fail outcomes and comparing the observed real-world rate with the resulting distribution~\cite{efron1994bootstrap}. Two-qubit analyses use Mann--Whitney comparisons of absolute deviation by pass/fail and Spearman correlations with operational deltas. Statistical significance is assessed at $\alpha=.05$ after correction where multiple comparisons are performed.

\section{Results} \label{sec:results}
\subsection{Semantic Translation Accuracy}
\begin{figure*}[!t]
    \centering
    \includegraphics[width=0.75\textwidth]{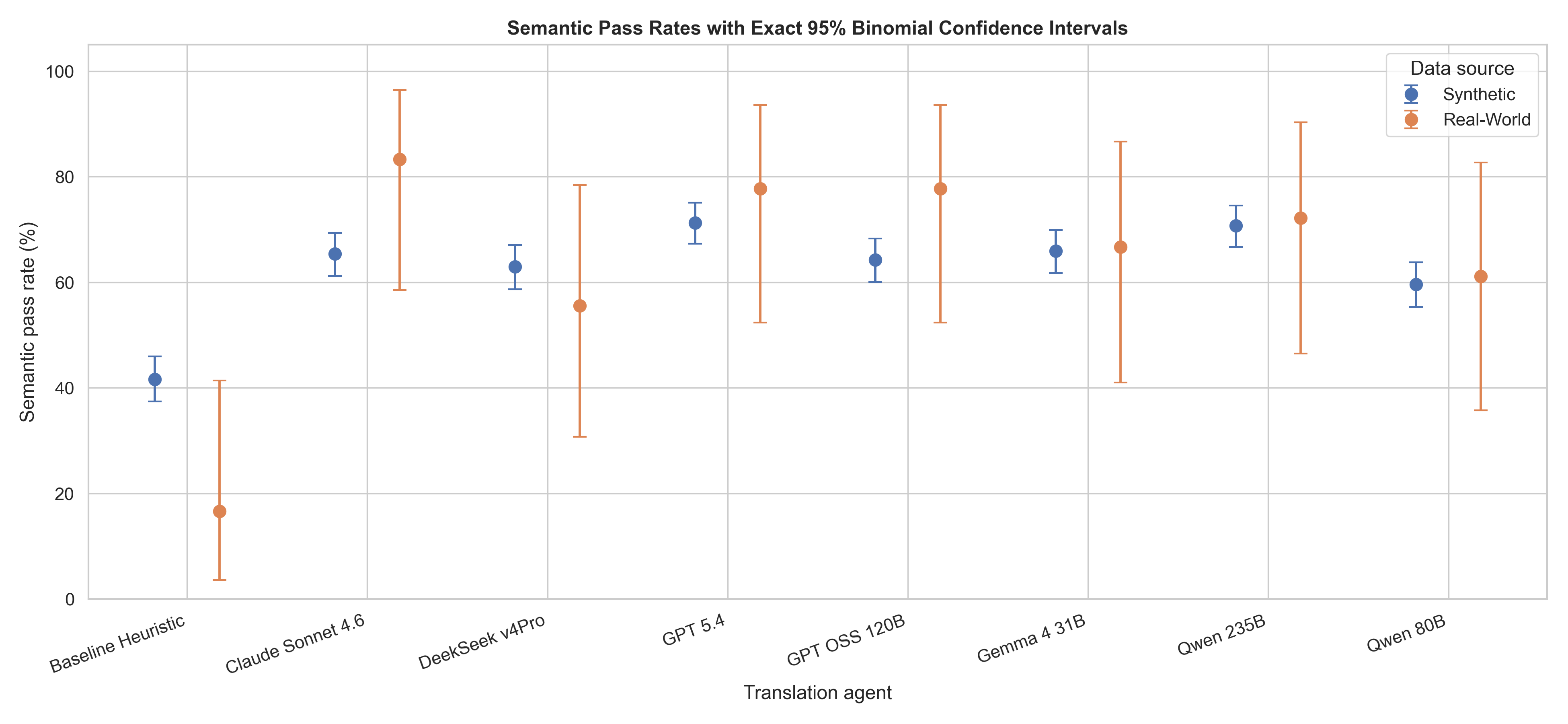}
    \caption{Semantic pass rates with Clopper--Pearson exact 95\% confidence intervals. Real-world intervals reflect substantial uncertainty from the 18-story holdout.}
    \label{fig:passci}
\end{figure*}

Table~\ref{tab:passci} reports semantic pass rates by model and data source. Every LLM agent outperformed the deterministic heuristic in both datasets. Claude Sonnet 4.6 achieved the highest real-world pass rate at 83.3\%, while GPT 5.4 and GPT OSS 120B each achieved 77.8\%.
Figure~\ref{fig:passci} reports point estimates and exact 95\% binomial confidence intervals. Synthetic estimates are precise because $N=540$, while the real-world intervals are wide because $N=18$.

\begin{table*}[!t]
\caption{Semantic pass rates and exact 95\% binomial confidence intervals.}
\label{tab:passci}
\centering
\scriptsize
\begin{tabular}{@{}lrrlrrl@{}}
\toprule
& \multicolumn{3}{c}{Synthetic ($N=540$)} & \multicolumn{3}{c}{Real-world ($N=18$)}\\
\cmidrule(lr){2-4}\cmidrule(lr){5-7}
Agent & Passes & SPR & 95\% CI & Passes & SPR & 95\% CI \\
\midrule
Baseline Heuristic & 225 & 41.7\% & [37.5, 46.0] & 3 & 16.7\% & [3.6, 41.4] \\
Claude Sonnet 4.6 & 353 & 65.4\% & [61.1, 69.4] & 15 & 83.3\% & [58.6, 96.4] \\
DeepSeek v4Pro & 340 & 63.0\% & [58.7, 67.0] & 10 & 55.6\% & [30.8, 78.5] \\
GPT 5.4 & 385 & 71.3\% & [67.3, 75.1] & 14 & 77.8\% & [52.4, 93.6] \\
GPT OSS 120B & 347 & 64.3\% & [60.1, 68.3] & 14 & 77.8\% & [52.4, 93.6] \\
Gemma 4 31B & 356 & 65.9\% & [61.8, 69.9] & 12 & 66.7\% & [41.0, 86.7] \\
Qwen 235B & 382 & 70.7\% & [66.7, 74.5] & 13 & 72.2\% & [46.5, 90.3] \\
Qwen 80B & 322 & 59.6\% & [55.4, 63.8] & 11 & 61.1\% & [35.7, 82.7] \\
\bottomrule
\end{tabular}
\end{table*}

\subsection{RQ1: Paired Model Comparisons}
Cochran's Q rejected equal pass probabilities on both the synthetic benchmark ($Q=216.41$, $df=7$, $p<10^{-42}$) and the exploratory real-world holdout ($Q=37.09$, $df=7$, $p<10^{-5}$). On the synthetic benchmark, every LLM significantly outperformed the heuristic after Holm correction. GPT 5.4 and Qwen 235B also significantly exceeded several other LLMs, although neither dominated every pairwise comparison.

On the real-world holdout, exact McNemar tests with Holm correction showed that the heuristic was significantly worse than Claude Sonnet 4.6 ($\Delta\spr=-66.7$ percentage points, adjusted $p=.0137$), GPT 5.4 ($\Delta\spr=-61.1$, adjusted $p=.0264$), and GPT OSS 120B ($\Delta\spr=-61.1$, adjusted $p=.0264$). No LLM-to-LLM pair remained significant after correction. Thus, the holdout supports the broad superiority of contextual LLM translation over the keyword baseline but does not support a definitive ranking among LLMs. The complete pairwise McNemar matrix, corrected and uncorrected, for both datasets is provided in the replication package's supplementary statistical tables rather than enumerated here.

\subsection{Safety-policy-related failures}
During error analysis, we manually reviewed the 13 Claude Sonnet 4.6 synthetic cases whose evaluation record returned \texttt{Failed.} rather than a scored explanation, and then traced those story IDs back to the seed prompts in \texttt{generate\_sample.py}. All 13 came from five hazard-themed synthetic seeds---volcanic aerosol dispersion, stratospheric greenhouse/geoengineering, antiviral molecule generation, viral-target protein-structure retrieval, and antibody--viral-spike docking---which together account for 50 of the 540 synthetic stories because each seed is crossed with the same ten intent templates. No real-world holdout story matched that same hazard-theme definition, so the real-world exclusion set size is $0/18$.

Under the study's strict protocol, Claude passed $0/13$ of this flagged synthetic subset, whereas the same stories were passed by the baseline heuristic in $7/13$ cases, DeepSeek v4Pro in $12/13$, Gemma 4 31B in $7/13$, GPT-5.4 in $12/13$, GPT-OSS 120B in $10/13$, Qwen 80B in $12/13$, and Qwen 235B in $10/13$ (Table~\ref{tab:hazard_sensitivity}). Excluding the 13 flagged synthetic stories from both numerator and denominator therefore raises Claude Sonnet 4.6's synthetic SPR from $353/540 = 65.4\%$ to $353/527 = 67.0\%$, while its real-world SPR remains $15/18 = 83.3\%$ because the real-world exclusion set is empty. This sensitivity analysis strengthens rather than reverses Claude's strict-protocol standing, but the effect is modest in magnitude and concentrated in a thematically narrow subset where refusal behavior, not semantic misinterpretation alone, appears to drive the score.

\begin{table}[t]
\caption{Hazard-themed subset sensitivity analysis.}
\label{tab:hazard_sensitivity}
\centering
\begin{adjustbox}{width=\columnwidth}
\begin{tabular}{lcc}
\toprule
Agent & Passes on 13 flagged synthetic stories & Subset SPR \\
\midrule
Baseline heuristic & $7/13$ & 53.8\% \\
Claude Sonnet 4.6 & $0/13$ & 0.0\% \\
DeepSeek v4Pro & $12/13$ & 92.3\% \\
Gemma 4 31B & $7/13$ & 53.8\% \\
GPT-5.4 & $12/13$ & 92.3\% \\
GPT-OSS 120B & $10/13$ & 76.9\% \\
Qwen 80B & $12/13$ & 92.3\% \\
Qwen 235B & $10/13$ & 76.9\% \\
\midrule
Claude Sonnet 4.6 rate & Numerator/denominator & SPR \\
\midrule
Strict-protocol synthetic SPR & $353/540$ & 65.4\% \\
Refusal-excluded synthetic SPR & $353/527$ & 67.0\% \\
Strict-protocol real-world SPR & $15/18$ & 83.3\% \\
Refusal-excluded real-world SPR & $15/18$ & 83.3\% \\
\bottomrule
\end{tabular}
\end{adjustbox}
\end{table}

\subsection{RQ2: Matched-Size Domain Shift}
Table~\ref{tab:shift} compares each real-world rate with 5,000 synthetic bootstrap samples of the same size ($N=18$). Only the heuristic fell outside its synthetic 95\% interval and reached a two-sided empirical $p<.05$. The LLM point estimates varied in direction, but none showed statistically detectable domain shift under the matched-size procedure.

\begin{table*}[!t]
\caption{Matched-size bootstrap domain-shift analysis.}
\label{tab:shift}
\centering
\scriptsize
\begin{tabular}{@{}lrrrrrl@{}}
\toprule
Agent & Synthetic mean & Synthetic 95\% interval & Real-world SPR & Difference & Two-sided $p$ & Interpretation \\
\midrule
Baseline Heuristic & 41.7\% & [22.2, 66.7] & 16.7\% & $-25.1$ pp & .0468 & Significant negative shift \\
Claude Sonnet 4.6 & 65.5\% & [44.4, 88.8] & 83.3\% & $+17.8$ pp & .1660 & No detectable shift \\
DeepSeek v4Pro & 62.9\% & [38.9, 83.3] & 55.6\% & $-7.4$ pp & .6647 & No detectable shift \\
GPT 5.4 & 71.2\% & [50.0, 88.9] & 77.8\% & $+6.5$ pp & .7495 & No detectable shift \\
GPT OSS 120B & 64.4\% & [44.4, 83.3] & 77.8\% & $+13.4$ pp & .3459 & No detectable shift \\
Gemma 4 31B & 66.2\% & [44.4, 83.3] & 66.7\% & $+0.5$ pp & 1.000 & No detectable shift \\
Qwen 235B & 70.8\% & [50.0, 88.9] & 72.2\% & $+1.4$ pp & 1.000 & No detectable shift \\
Qwen 80B & 59.5\% & [38.9, 83.3] & 61.1\% & $+1.6$ pp & 1.000 & No detectable shift \\
\bottomrule
\end{tabular}
\end{table*}

\subsection{RQ3: Intent-Dependent Operational Outcomes}
\begin{figure*}[!t]
\centering
\begin{subfigure}[t]{0.48\textwidth}
\centering
\includegraphics[width=\linewidth]{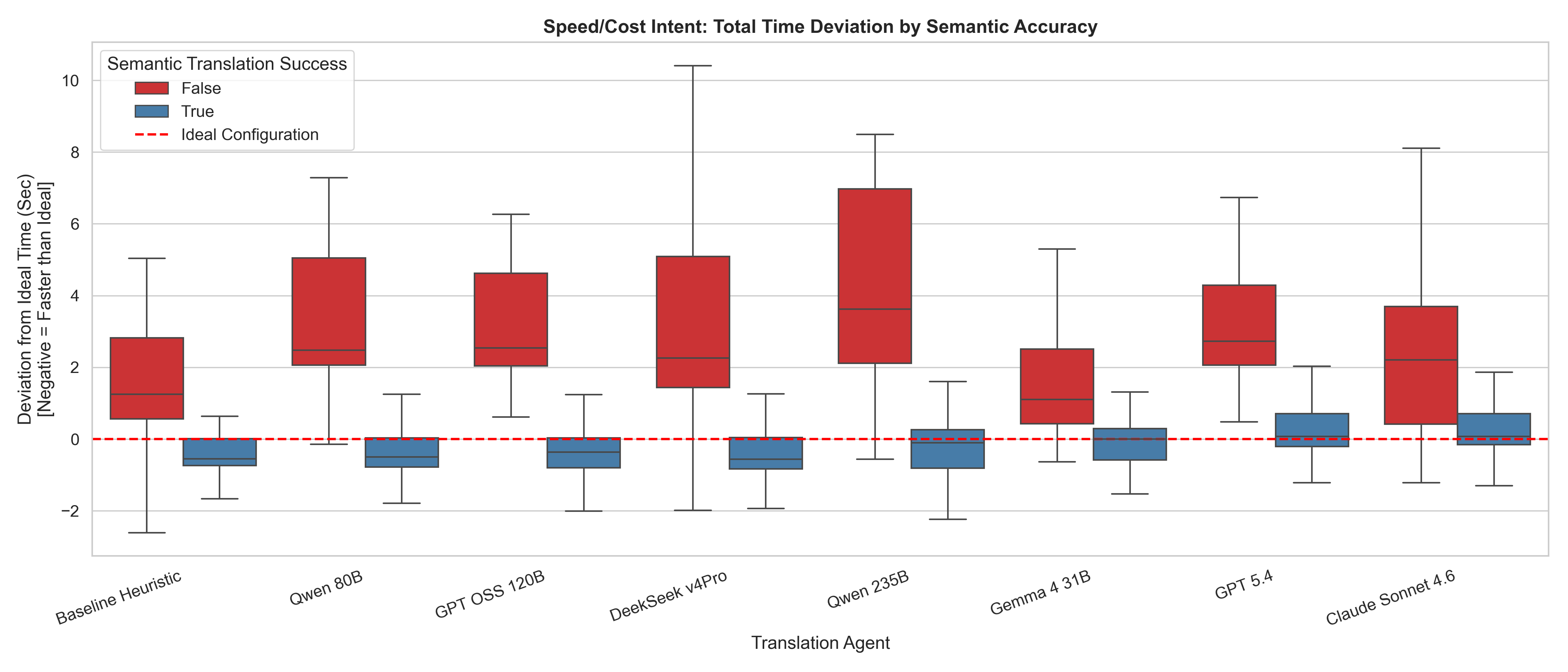}
\caption{Time deviation by semantic outcome.}
\end{subfigure}\hfill
\begin{subfigure}[t]{0.48\textwidth}
\centering
\includegraphics[width=\linewidth]{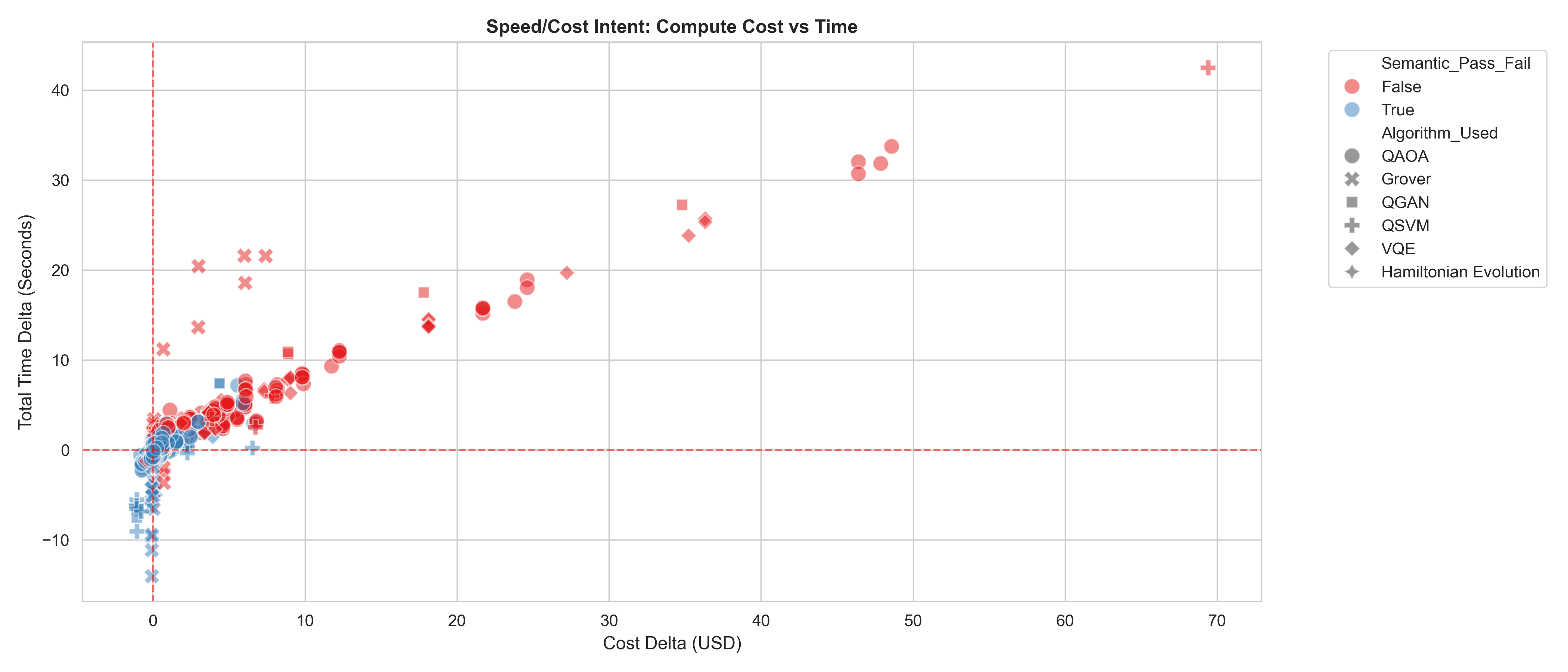}
\caption{Cost versus time deviation.}
\end{subfigure}
\caption{Operational penalties for Speed/Cost requirements. Failed translations more often occupy slower and more expensive regions.}
\label{fig:speedpenalty}
\end{figure*}

\begin{figure*}[!t]
\centering
\begin{subfigure}[t]{0.48\textwidth}
\centering
\includegraphics[width=\linewidth]{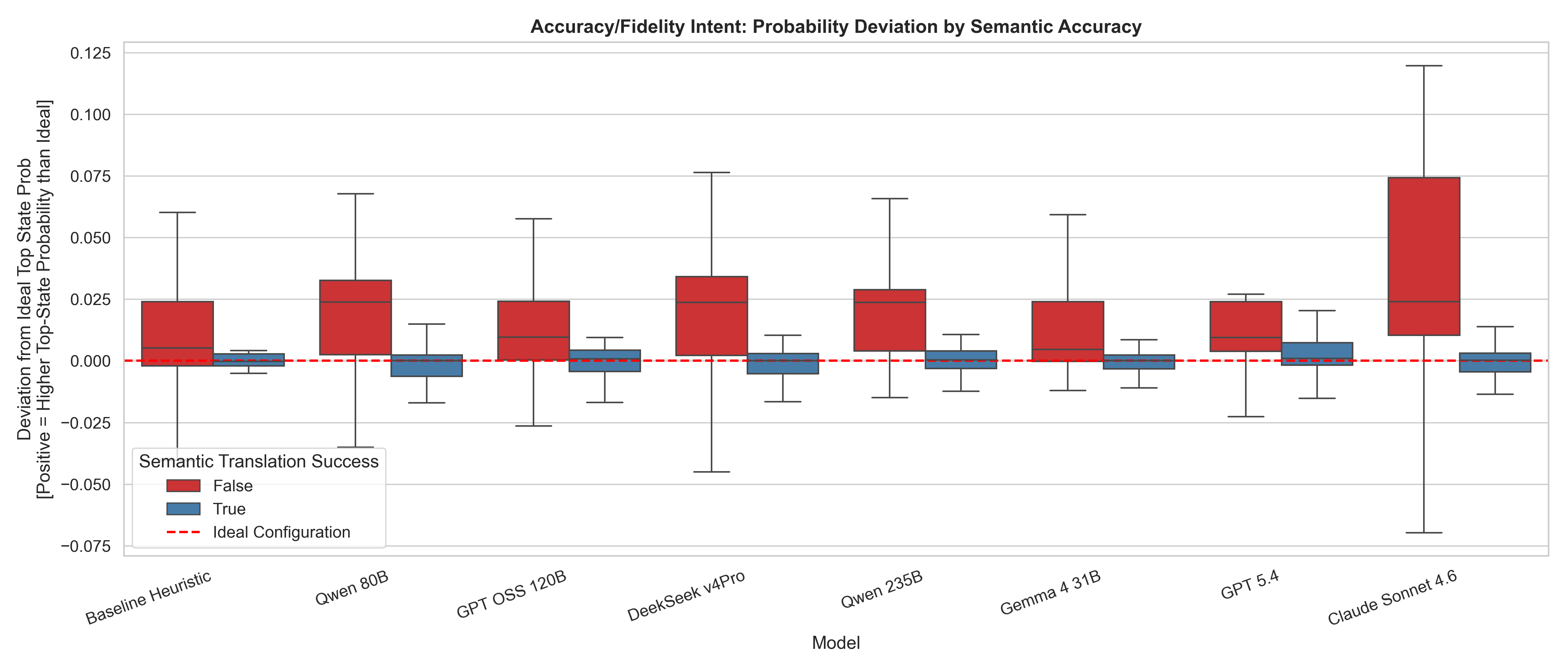}
\caption{Top-state probability deviation by semantic outcome.}
\end{subfigure}\hfill
\begin{subfigure}[t]{0.48\textwidth}
\centering
\includegraphics[width=\linewidth]{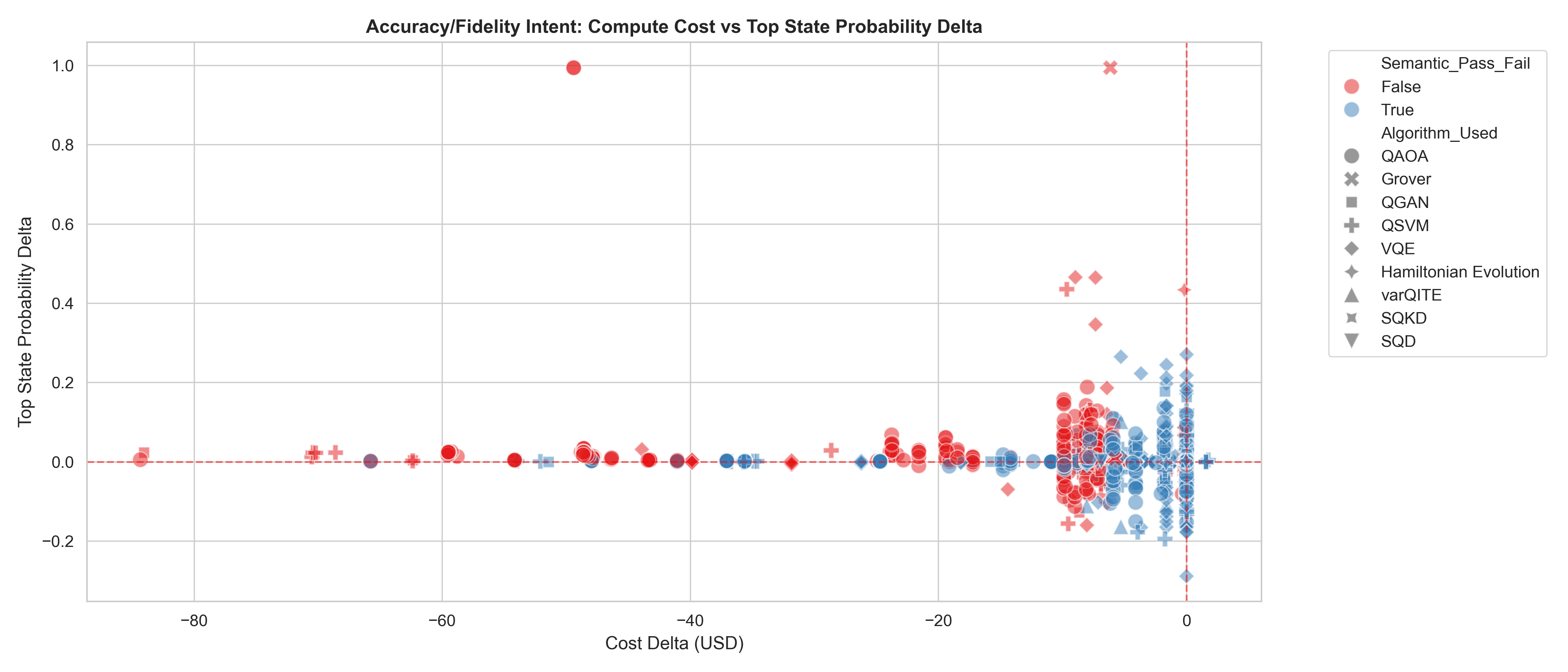}
\caption{Cost versus probability deviation.}
\end{subfigure}
\caption{Accuracy/Fidelity results. Top-state probability does not exhibit a consistent semantic-success pattern.}
\label{fig:probpenalty}
\end{figure*}

For successful synthetic translations, Speed/Cost stories had significantly lower execution time and cost than Accuracy/Fidelity stories for every model (Figure~\ref{fig:speedpenalty}). The corresponding $\aonetwo$ values indicated large separation. The top-state probability hypothesis was not supported for any model (Figure~\ref{fig:probpenalty}): $p$-values ranged from .522 to .960 and effects were negligible or small.

Paired Friedman tests further showed model-dependent differences on the synthetic benchmark for absolute Speed/Cost time deviation ($\chi^2_F=85.67$, $p<10^{-15}$), cost deviation ($\chi^2_F=202.73$, $p<10^{-39}$), and Accuracy/Fidelity probability deviation ($\chi^2_F=23.50$, $p=.00139$). The exploratory real-world probability comparison was not significant ($\chi^2_F=5.48$, $p=.602$). Pairwise Wilcoxon tests with Holm correction identify where those omnibus differences occur. For example, on Speed/Cost stories, Claude Sonnet 4.6 had lower absolute time deviation than Qwen 235B (median $0.665$ s vs.\ $0.950$ s, Holm-adjusted $p=2.47\times 10^{-15}$, paired rank-biserial $|r_{rb}|=.656$), and DeepSeek v4Pro had lower absolute cost deviation than Qwen 235B (median \$0.0196 vs.\ \$0.1338, Holm-adjusted $p=6.69\times 10^{-12}$, $|r_{rb}|=.810$). The detailed pairwise tables remain in the supplementary materials due to the high number of comparisons.

\subsection{Secondary Analysis: Two-Qubit Gate Exposure}
Two-qubit gate deviation was zero for the median observation in nearly every model/domain group. Samples densely clustered at the ideal zero line with a small number of large positive and negative outliers. Thus, parameter translation usually preserved entangling-gate burden, but occasional configurations substantially changed routing or decomposition.

The synthetic paired Friedman test detected model differences in absolute two-qubit deviation ($\chi^2_F=29.35$, $df=6$, $p=.000052$), but pairwise median differences were zero and appear driven by the frequency and magnitude of outliers. The real-world omnibus test was not significant ($\chi^2_F=6.00$, $p=.423$). Failure-versus-pass Mann--Whitney comparisons were generally non-significant or negligible in magnitude. Isolated real-world findings for GPT OSS 120B, GPT 5.4, and Claude Sonnet 4.6 are treated as exploratory because the failure groups contained only three or four observations.

\begin{figure*}[!t]
    \centering
    \includegraphics[width=0.6\textwidth]{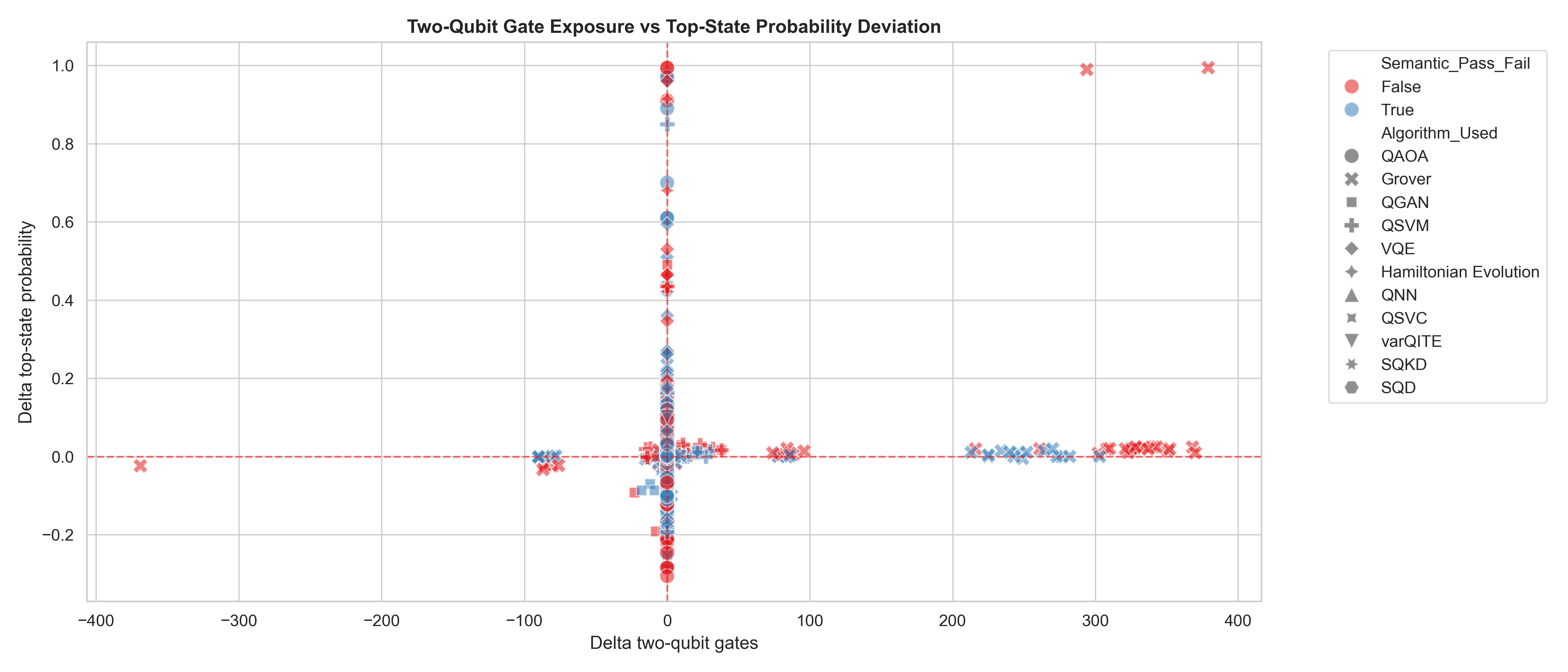}
    \caption{Relationship between two-qubit gate deviation and top-state probability deviation. The vertical concentration at zero illustrates that probability changes frequently occur without a change in two-qubit gate count.}
    \label{fig:g2prob}
\end{figure*}

Across all synthetic LLM observations, $\gdelta$ had only a negligible positive association with top-state probability deviation ($\rho=.085$, $p<10^{-6}$). The real-world association was not significant ($\rho=.148$, $p=.099$). Figure~\ref{fig:g2prob} confirms that large probability deviations often occur even when gate-count deviation is zero. Therefore, two-qubit gate count is informative as an exposure and outlier metric, but it does not explain the fidelity-proxy behavior in this experiment.

The approximate $\epsilon_{2q}=.01$ noise-risk transform preserves the same qualitative pattern and is supplied as a sensitivity visualization rather than a calibrated hardware-fidelity estimate.

\begin{table}[!t]
\caption{Summary of findings by research question.}
\label{tab:summary}
\begin{tabularx}{\columnwidth}{@{}lY@{}}
\toprule
RQ & Finding \\
\midrule
RQ1 & LLMs outperform the heuristic baseline. Real-world LLM ranking remains uncertain. \\
RQ2 & Only the heuristic baseline shows statistically detectable negative domain shift. \\
RQ3 & Time and cost hypotheses are supported but top-state probability is not. Two-qubit burden reveals rare structural outliers but weak fidelity-proxy association. \\
\bottomrule
\end{tabularx}
\end{table}

\section{Discussion} \label{sec:futurework}
\textbf{\textit{Principal Findings.}}
Table~\ref{tab:summary} summarizes the findings for each research question. The results support LLM-based semantic middleware over deterministic keyword mapping. All LLMs exceeded the heuristic descriptively, and paired tests confirmed broad differences on both datasets. However, because the real-world sample size was small, the confidence intervals were too wide to establish a definitive, statistically significant ranking among the top-performing LLMs. Therefore, the primary conclusion is that contextual AI is broadly more robust and reliable than rigid baseline methods, rather than one specific model being strictly superior.

\textbf{\textit{Synthetic and Real-World Evidence Play Different Roles.}}
The study deliberately utilizes two datasets for distinct methodological purposes. The large synthetic dataset provides the statistical power required for controlled model comparisons, while the real-world holdout offers practical, ecological validation. The baseline heuristic exhibits a statistically detectable decline in performance (domain shift) when applied to real-world developer inputs. Consequently, broader claims about how well LLMs generalize to real-world quantum development should be restrained until larger datasets are evaluated.

\textbf{Operational Misalignment.}
The Speed/Cost analyses show why semantic correctness matters beyond JSON validity. Incorrectly high shot counts can make a configuration substantially slower and more expensive while still executing successfully. This establishes NFR preservation as an operational quality concern and motivates simulation previews, guardrails, and human review before costly execution.

\textbf{\textit{Interpreting the Fidelity Findings.}}
Top-state probability did not support the Accuracy/Fidelity hypothesis. This metric measures sampling concentration, not statevector fidelity, energy error, approximation ratio, or task accuracy. Moreover, shot count changes sampling variance but not the underlying accuracy of the quantum state. For variational algorithms simulation, random parameters introduce additional variability. The null result should therefore be interpreted as a limitation of the chosen proxy metric rather than proof that accuracy-oriented configurations cannot improve task quality.

\textbf{\textit{Two-Qubit Gate Exposure.}}
The secondary analysis adds an important structural perspective. Predicted and ideal circuits usually had identical two-qubit gate counts, so translation choices often changed sampling effort without changing entangling-gate burden. Rare outliers show that optimization and routing can still generate substantial structural penalties. However, the negligible association with top-state probability confirms that raw two-qubit count is not a direct fidelity measure. Future work should combine backend-specific gate error rates, circuit depth, critical-path scheduling, and algorithm-specific output metrics.

\textbf{\textit{Implications for QSE Tools.}}
A practical system should preserve the source requirement, infer an explicit macro-intent, emit schema-validated parameters, enforce strict parameter boundaries, estimate time/cost/circuit consequences, and request human clarification when uncertainty is high. The LLM should be treated as a recommendation component rather than an autonomous execution authority.

None of these design implications are quantum-specific: any NL-to-structured-configuration middleware, classical or quantum, faces the same need to preserve intent, bound outputs, and validate against execution consequences rather than textual similarity to a reference answer. QSE simply makes the cost of an unbounded or unvalidated translation immediately quantifiable, which is why we used it as the evaluation setting motivated in Section~\ref{sec:introduction}. The methodology, not just the conclusion, should transfer to other RE-to-configuration translation problems where execution behavior, rather than surface output, is the right criterion for correctness.

The two-parameter scope adopted here (Section~\ref{sec:background}) was chosen for cross-backend tractability, not because \texttt{shots} and \texttt{optimization\_level} exhaust the configuration surface. The same correctness-to-cost methodology could extend to backend selection, error mitigation method, and qubit layout, but those parameters are backend- and hardware-specific and would require a per-backend ground-truth rubric and simulation model rather than the single shared rubric used here. We view this as a natural next step once the present, more constrained translation task is well understood.

\section{Threats to Validity} \label{sec:threats}
\textbf{\textit{Construct Validity.}}
Semantic Pass Rate compresses continuous parameters like shots into broad intent categories, improving interpretability but treating different in-range values as equivalent. Top-state probability is not full fidelity, the two-qubit noise exposure uses a fixed illustrative error rate rather than calibration data from a physical processor, and \texttt{FakeBrisbane} itself is a static noise-model snapshot rather than live, continuously updated hardware calibration, so absolute fidelity values may not track the present-day device. The rubric also only checks \texttt{shots} and \texttt{optimization\_level} against categorical ranges, so a ``semantic pass'' does not certify full NFR comprehension: a translation can land in the correct bucket by chance, a generic default, or partial understanding. Reported pass rates should therefore be read as a lower bound on translation reliability rather than a direct measure of NFR understanding.

\textbf{\textit{Internal Validity.}}
Using an LLM for both translation and judging risks shared-model or shared-prompt bias, mitigated but not eliminated by the explicit rubric; variational parameters and transpilation add stochastic variation that median aggregation only partly absorbs, and future work should involve independent human adjudication. A related threat is that the 540 synthetic stories are generated by an LLM from author-authored templates, judged by an LLM, against ground truth defined by the same template-to-category mapping, forming a closed loop in which generator regularities could inflate pass rates independent of judge bias (Section~\ref{sec:methodology}); the real-world holdout, despite its small size, is the main check against this. Finally, differential provider safety behavior is a threat: some Claude Sonnet 4.6 failures were associated with hazard-related stories that may have triggered refusal despite benign intent, and such cases were counted as failures even though they conflate translation accuracy with moderation policy.

\textbf{\textit{Conclusion Validity.}}
All models are evaluated on the same stories, so paired tests (Cochran's Q, exact McNemar, Friedman, Wilcoxon with Holm correction) are used throughout. The two-qubit distributions are highly skewed toward zero, so significant omnibus results can reflect outlier frequency rather than median shifts; we therefore report effect sizes and avoid equating significance with practical importance.

\textbf{\textit{External Validity.}}
The synthetic benchmark may inherit generator-language regularities from its templates, and the 18-story real-world holdout may not represent all quantum development contexts; a larger, more diverse collection of real-world projects is needed before these results generalize.

\textbf{\textit{Reliability and Reproducibility.}}
Model APIs, provider versions, and decoding behavior can change, so prompt templates, exact model identifiers, code, package versions, random seeds, raw outputs, and statistical tables should be archived.

\section{Conclusion} \label{sec:conclusion}
This study evaluated LLMs as semantic middleware for translating natural-language NFRs into Qiskit shot and transpiler-optimization settings. Across 540 synthetic and 18 real-world stories, LLMs substantially outperformed a deterministic keyword baseline. Paired testing confirmed model differences, while exact intervals showed that the small real-world sample cannot support a precise ranking among the strongest LLMs. Matched-size bootstrapping detected a negative domain shift only for the heuristic.

The operational results demonstrate that semantic errors can violate speed and cost goals even when a circuit executes correctly. In contrast, top-state probability did not validate the proposed fidelity hypothesis. The two-qubit analysis showed that predicted and ideal configurations usually preserve entangling-gate count, with rare but large outliers and only a weak relationship to top-state probability.

LLM-assisted configuration is promising but should remain bounded and reviewable. Future work should enlarge the real-world benchmark, use human or multi-judge adjudication, evaluate repeated stochastic LLM runs, incorporate backend-calibrated error data, and replace generic top-state probability with algorithm-specific quality measures such as VQE energy error, QAOA approximation ratio, Grover marked-state success, and QML predictive accuracy. 




\appendices
\section{Prompt Template Snapshots}\label{app:prompts}
The following listings reproduce the core prompt structure used in the experiment. Provider wrappers and Pydantic-generated schema text are omitted only where they are automatically generated; the semantic instructions are retained.

\begin{lstlisting}[caption={Synthetic user-story generator prompt snapshot.},label={lst:syntheticprompt}]
SYSTEM:
You are a quantum software engineer and expert enterprise architect.
Your task is to generate synthetic data.

IMPORTANT:
- Output a valid JSON object containing actual generated values.
- Do not output schema properties or descriptions.
- Do not output Markdown blocks.
- Generate realistic, highly diverse business language based on the
  requested domain and constraint.

USER:
Generate one user story focusing on {domain}.
The story must imply a strict need to {constraint}.
\end{lstlisting}

\begin{lstlisting}[caption={Translation-agent prompt snapshot.},label={lst:translationprompt}]
SYSTEM:
You are a middleware agent translating classical user stories into
Qiskit parameters. Return only valid JSON matching the required schema.

USER:
Extract the implicit constraints from this story and predict the
parameters:
{user_story}
\end{lstlisting}

\begin{lstlisting}[caption={Rubric-driven judge prompt snapshot.},label={lst:judgeprompt}]
SYSTEM:
You are an expert Quantum Software Engineering evaluator grading an
AI Translation Agent. Compare the predicted parameters with the ideal
ground truth using semantic intent.

RUBRIC:
1. SPEED/COST (ideal shots approximately 10-500)
   - score 5 if predicted_shots <= 1000; otherwise score 0.
2. ACCURACY/FIDELITY (ideal shots approximately 4000-10000)
   - score 5 if predicted_shots >= 2000; otherwise score 0.
3. BALANCED/SCALING (ideal shots approximately 1024-2048)
   - score 5 if 500 <= predicted_shots <= 4000; otherwise score 0.
4. OPTIMIZATION LEVEL (0-3)
   - score 5 if predicted level is within +/-1 of the ideal;
     otherwise score 0.

Set pass_fail=true only if BOTH the shot score and the optimization
score equal 5.
Output only valid JSON. Keep the explanation to at most three sentences.

USER:
Constraint Requirement: {implicit_nfrs}
User Story: {user_story}
Ideal: Shots={ideal_shots}, Opt={ideal_optimization}
Predicted: Shots={predicted_shots}, Opt={predicted_optimization}
\end{lstlisting}

\bibliographystyle{IEEEtran}
\bibliography{references}

\end{document}